# Structural transitions, magnetic properties, and electronic structures of Co(Fe)-doped MnNiSi compounds

Y. Li,[1,2] Z. Y. Wei,[1] E. K. Liu,[1,a)] G. D. Liu,[2] S. G. Wang,[1] W. H. Wang,[1] and G. H. Wu[1]

[1] *State Key Laboratory for Magnetism, Beijing National Laboratory for Condensed Matter Physics, Institute of Physics, Chinese Academy of Sciences, Beijing 100190, China*

[2] *School of Material Science and Engineering, Hebei University of Technology, Tianjin 300130, China*

ABSTRACT

The structural transitions, magnetic properties, and electronic structures of Co(Fe)-doped MnNiSi compounds are investigated by x-ray powder diffraction (XRD), differential scanning calorimetry (DSC), magnetic measurements, and first-principles calculations. Results indicate that all samples undergo a martensitic transition from the $Ni_2In$-type parent phase to TiNiSi-type orthorhombic phase at high temperatures. The substitution of Co(Fe) for Mn in $Mn_{1-x}Co_xNiSi$ ($x$ = 0.2, 0.3, 0.4) and $Mn_{1-y}Fe_yNiSi$ ($y$ = 0.26, 0.30, 0.36, 0.46, 0.55) samples decreases the structural transition temperature ($T_t$) and Curie temperature of martensite ($T_C^M$). The martensite phases show a typical ferromagnetic behavior with saturation field ($H_S$) being basically unchanged with increasing Co(Fe) content, while the saturation magnetization ($M_S$) shows a decreasing tendency. The theoretically calculated moments are in good agreement with the experimentally measured results. The orbital hybridizations between different 3$d$ elements are analyzed from the distribution of density of states.

a) E-mail: ekliu@iphy.ac.cn

## I. INTRODUCTION

As new magnetic phase-transition functional materials, the $Ni_2In$-type hexagonal MM'X (M, M' = transition metals, X = carbon or boron group elements) intermetallic compounds have been widely studied due to the ferromagnetic (FM) shape memory behavior, giant magnetocaloric and magnetoelastic effects for potential applications.[1-6] In recent years, several effective methods such as the application of pressure,[7] chemical substitution,[8-11] vacancy introduction[1-3] and isostructural alloying,[5,12-14] have been used to tune the first-order martensitic transitions (MT) for magnetoresponsive effects. In previous works, it has been reported that MnNiGe[5] shows first-order MT with $T_t$ ~470 K between paramagnetic (PM) $Ni_2In$-type parent phase and PM TiNiSi-type martensitic phase, with a Néel temperature $T_N$ of ~346 K for the spiral antiferromagnetic (AFM) martensite. Doping Co(Fe) in MnNiGe[5,13,15] can lead to tunable PM/FM-type MTs in the temperature range from room to cryogenic temperatures, achieving a simultaneous manipulation on the phase transition and magnetic couplings. Giant magnetocaloric effects of over -30 J $kg^{-1}$ $K^{-1}$ at a field change of 50 kOe were estimated, which puts these compounds into promising giant magnetocaloric materials. Further studies on new materials with desired properties are still in progress.

Stoichiometric MiNiSi[16,17] MM'X compound undergoes a MT at 1210 K from a hexagonal phase ($Ni_2In$-type, space group $P6_3/mmc$) to orthorhombic phase (TiNiSi-type, space group Pnma). The martensite phase is a typical ferromagnet with a high Curie temperature of martensite ($T_C^M$) at 615 K. By alloying MnNiSi with MnNiGe, the structural and magnetic properties were studied based on the substitution of main-group elements.[16] Until now, the influence of transition-metal Co(Fe) doping on the structural and magnetic properties of MnNiSi alloys is still unclear. In this work, combining experimental and theoretical results, we study Co(Fe)-doped MnNiSi compound and discuss structure transitions, magnetic properties, and electronic structures in both $Mn_{1-x}Co_xNiSi$ and $Mn_{1-y}Fe_yNiSi$ systems.

## II. EXPERIMENT

Polycrystalline ingots of $Mn_{1-x}Co_xNiSi$ ($x$ = 0.2, 0.3, 0.4) and $Mn_{1-y}Fe_yNiSi$ ($y$ = 0.26, 0.30, 0.36, 0.46, 0.55) alloys were prepared by arc melting high-purity metals under argon atmosphere. The as-cast ingots were annealed at 1123 K in an evacuated quartz tube for five days and then cooled slowly to room temperature. The structural characterizations were performed by powder x-ray diffraction (XRD) with Cu-$K\alpha$ radiation at room temperature. The differential scanning calorimetry (DSC) with magnet-assisted thermogravimetric analysis (TGA) was used to detect the structural-transition temperatures and Curie temperatures. Two permanent magnets were mounted on both sides of the sample chamber of the DSC with a proper vertical height to the sample position. With the aid of the magnets, a measured change in sample mass can be detected by the TGA when the sample undergoes a magnetic transition between PM and FM states, since the magnetic attractive force will appear or disappear between the magnets and the magnetized samples. The rate of heating/cooling was set as 10 K/min. The magnetizations as a function of applied field at 5 K in fields up to 50 kOe were performed on a superconducting quantum interference device (SQUID). The CASTEP codes[18] using the pseudopotential method with plane-wave-basis set based on the density-functional theory was applied to calculate the magnetic moments and density of states of the studied alloys. The exchange correlation energy for the structural relaxations and the electronic structures was treated under the generalized-gradient approximation (GGA)[19] and the local-density approximation (LDA),[20] respectively. The plane-wave cutoff energy of 400 eV and 120 (8×10×6) $k$ points in the irreducible Brillouin zone were used to insure a good convergence of the total energy.

III. RESULTS AND DISCUSSION

XRD patterns of $Mn_{1-x}Co_xNiSi$ ($x$ = 0.2, 0.3, 0.4) and $Mn_{1-y}Fe_yNiSi$ ($y$ = 0.26, 0.30, 0.36, 0.46, 0.55) at room temperature are shown in Figs. 1(a) and (b). As mentioned above, the stoichiometric MnNiSi compound shows a TiNiSi-type orthorhombic structure at room temperature. In Fig. 1(a), after Co atoms were introduced into $Mn_{1-x}Co_xNiSi$, the series of samples show a TiNiSi-type orthorhombic structure, indicating that their $T_t$ are higher than room temperature. At the same time, the residual $Ni_2In$-type hexagonal parent phase can also be observed. In Fig. 1(b), after Fe atoms were introduced into $Mn_{1-y}Fe_yNiSi$, the similar room-temperature orthorhombic structure with a little residual parent phase is also observed. The experimental lattice constants of $Mn_{1-x}Co_xNiSi$ and $Mn_{1-y}Fe_yNiSi$ are given in Fig. 2 and Table I. We can see from Fig. 2(a) that, for both alloy systems, doping Co(Fe) in MnNiSi declines the lattice parameter $a$ in orthorhombic structure and imposes little influence on the parameters of $b$ and $c$. In Co(Fe)-doped MnNiSi compounds, the atomic radius of Co(Fe) are smaller than that of Mn atom, and the hexatomic rings consisted of alternate Ni and Si atoms keep unchanged.[21] After the Co(Fe) atoms were substituted for Mn atoms, the distance between Mn atom and Ni-Si ring will decrease due to the formation of the chemical bonding, which results in a decrease in $a$ axis of orthorhombic structure.

The Curie magnetic transitions and first-order MTs of $Mn_{1-x}Co_xNiSi$ ($x$ = 0.2, 0.3, 0.4) and $Mn_{1-y}Fe_yNiSi$ ($y$ = 0.26, 0.30, 0.36, 0.46, 0.55) samples were measured by DSC upon cooling. The data are listed in Table I and plotted in Figs. 3(a) and (b). As shown in the inset of Fig. 3(b), the sharp exothermic peak at 955 K without a mass change indicates the first-order MT from PM $Ni_2In$-type hexagonal structure to PM TiNiSi-type orthorhombic one in $Mn_{0.74}Fe_{0.26}NiSi$ sample ($y$ = 0.26). From Figs. 3(a) and (b), $T_t$ of both alloy systems decreases with increasing Co(Fe) content. It can be further seen that to some extent $T_t$ of $Mn_{1-y}Fe_yNiSi$ is decreased more rapidly than that of $Mn_{1-x}Co_xNiSi$, which indicates that Fe shows a stronger effect on the phase stabilization of the parent phase. The Curie temperature of martensite ($T_C^M$) was determined by magnet-assisted DSC + TGA. For $Mn_{0.74}Fe_{0.26}NiSi$ sample, the exothermic peak at 583 K corresponds to a mass reduction during cooling, which means that the sample undergoes the Curie transition and the upward magnetic force was produced between the magnetized samples and the permanent magnets. With increasing Co(Fe) content, $T_C^M$ of the samples also decreases, which indicates the substitution of Co(Fe) for Mn will destroy the exchange interactions between Mn-Mn atoms in MnNiSi. From Fig. 3, we can see the martensitic structural transition of Co(Fe)-doped MnNiSi happens above the Curie temperatures of martensite phase in the present Co(Fe) doping levels. At room temperature, the ferromagnetic martensite phase is obtained with Curie temperatures above 400 K.

In order to analyze the magnetic behavior, the magnetization curves of $Mn_{1-x}Co_xNiSi$ ($x$ = 0.2, 0.3, 0.4) and $Mn_{1-y}Fe_yNiSi$ ($y$ = 0.26, 0.30, 0.36, 0.46, 0.55) samples were further measured at 5 K as a function of applied magnetic field up to 50 kOe, as shown in Fig. 4. As seen in Fig. 4(a), the three samples of $Mn_{1-x}Co_xNiSi$ are all in martensite structure and show a typical ferromagnetically magnetizating behavior. The saturation magnetization ($M_S$) decreases significantly with increasing Co content. The saturation field ($H_S$) of martensite phase shows a slight increase, indicating Co atoms to some extent destroy the strong coupling between Mn moments. For $Mn_{1-y}Fe_yNiSi$, the magnetization curves indicate a similar FM behavior, as shown in Fig. 4(c). In contrast, the $M_S$ shows a slight decrease with increasing Fe content. For this system, $H_S$ basically keeps unchanged with increasing Fe content.

At the same time, the total and atomic magnetic moments of some Co(Fe)-doped MnNiSi were calculated by first principles method, as shown in Figs. 4(b) and (d) and listed in Table II. The magnetic moment of MnNiSi (2.70 $\mu_B$) is highly accordant with the reported experimental value (2.62 $\mu_B$).[22] Besides, the calculated total moments for both alloy systems (green pentagrams ★ in Figs. 4(b) and (d)) show good coherence with the experimental results in this study. According to the calculated results, the atomic moments of Co and Mn are ~ 0.5 and ~2.5 $\mu_B$, respectively, during the substitution in $Mn_{1-x}Co_xNiSi$. Replacing the large-moment Mn atoms with small-moment Co atoms results in a reasonable decrease in the total moment. In $Mn_{1-y}Fe_yNiSi$, the slight decrease is attributed to the nearly same moment values of Fe (~1.6 $\mu_B$) and Mn (~2.4 $\mu_B$).

To further understand the electronic structures of the studied samples, we calculated the density of state (DOS) of Co(Fe)-doped MnNiSi. Figure 5 shows the total and the partial DOS for $Mn_{0.75}Fe_{0.25}NiSi$. The hybridization peaks appear at -1.5 and -1 eV for spin-up state, while -0.77 eV and around Fermi level for spin down state. There exists strong d-d orbit hybridizations between Mn and Fe atoms since two atoms share the same site and have direct chemical bonding. Mn/Fe atoms make the main contribution to the DOS distribution around Fermi level. In contrast, Ni shows a weaker hybridization strength with Mn/Fe, with lower-density peaks at the above energy levels, especially for the spin-up state. Instead, the high-density peaks of Ni locate at lower energy range between -5 and -1.5 eV, where the *p-d* orbit hybridizations between Si and Ni atoms are observed clearly in both spin-up and spin-down states, corresponding to the strong covalent bonding between the nearest-neighbor Ni and Si atoms in MnNiSi compound.[16] In particular, a strong *p-d* hybridization can be observed around -4 eV between Ni and Si atoms in both spin states. Furthermore, it can be seen that for Mn atom the spin-down DOS mainly locate above the Fermi level, while the spin-up states locate below the Fermi level. This results in a large spin splitting in two spin states and a large net magnetic moment (2.36 $\mu_B$, see Table II) on Mn atom. Compared with Mn atom, Fe atom shows a relatively weak spin splitting with more spin-down states being occupied. Fe atom thus carries a smaller moment (1.60 $\mu_B$, see Table II). Nevertheless, a great similarity of DOS distributions between Mn and Fe atoms can be observed. For Ni atom, since both spin states are almost completely occupied due to the *p-d* hybridizations at deep energies, a very small moment (0.22 $\mu_B$, see Table II) is observed. From the DOS distribution of $Mn_{0.75}Fe_{0.25}NiSi$, it can be seen that Ni-Si bonding serves mainly as covalent networks with small magnetic moments, while Mn/Fe atoms surrounded by covalent networks carry the main magnetic moments and determine the electronic structure around Fermi level.

## IV. CONCLUSIONS

The structural transitions, magnetic properties, and electronic structures of $Mn_{1-x}Co_xNiSi$ ($x$ = 0.2, 0.3, 0.4) and $Mn_{1-y}Fe_yNiSi$ ($y$ = 0.26, 0.30, 0.36, 0.46, 0.55) are studied in this work. All samples show ferromagnetic TiNiSi-type orthorhombic structures at room temperature. $T_t$ and $T_C^M$ of both alloy systems decrease with increasing Co(Fe) content. Compared with Co, Fe shows a stronger effect on the phase stabilization of the parent phase. The calculated magnetic moments provide a good explanation on the magnitude and the decrease tendency of the saturation magnetization measured in experiments. Surrounded by the Ni-Si covalent networks, Mn/Fe atoms carry the main magnetic moments and largely influence the electronic structures around Fermi level.

## ACKNOWLEDGEMENTS

This work was supported by National Natural Science Foundation of China (51301195 and 51431009), the National Basic Research Program of China (973 Program: 2012CB619405), Beijing Municipal Science & Technology Commission (Z141100004214004) and Youth Innovation Promotion Association of Chinese Academic of Sciences.

Tables

Table I. The experimental lattice constants $a$, $b$ and $c$, structural transition temperature ($T_t$), Curie temperature of martensite ($T_C^M$), saturation magnetization ($M_S$), and saturation field ($H_S$) for MnNiSi, Mn$_{1-x}$Co$_x$NiSi ($x$ = 0.2, 0.3, 0.4) and Mn$_{1-y}$Fe$_y$NiSi ($y$ = 0.26, 0.30, 0.36, 0.46, 0.55) compounds.

| | MnNiSi* | Mn$_{1-x}$Co$_x$NiSi ($x$) | | | Mn$_{1-y}$Fe$_y$NiSi ($y$) | | | | |
|---|---|---|---|---|---|---|---|---|---|
| | | 0.2 | 0.3 | 0.4 | 0.26 | 0.30 | 0.36 | 0.46 | 0.55 |
| $a$ (Å) | 5.901 | 6.0518 | 6.0350 | 6.0271 | 6.0549 | 6.0518 | 6.0499 | 6.0442 | 6.0421 |
| $b$ (Å) | 3.606 | 3.7541 | 3.7567 | 3.7580 | 3.8228 | 3.7864 | 3.7092 | 3.7633 | 7.7668 |
| $c$ (Å) | 6.902 | 7.0880 | 7.0924 | 7.0922 | 7.0703 | 7.0400 | 7.0813 | 7.0949 | 7.0830 |
| $T_t$ (K) | 1210 | 1091 | 997 | 839 | 955 | 932 | 866 | 648 | 554 |
| $T_C^M$ (K) | 622 | 592 | 560 | 541 | 583 | 580 | 552 | 479 | 438 |
| $M_S$ ($\mu_B$/f.u.) | 2.62 | 2.2767 | 1.9902 | 1.6985 | 2.3829 | 2.2505 | 2.2860 | 2.0932 | 2.1285 |
| $H_S$ (kOe) | - | 1.76 | 1.99 | 2.17 | 1.47 | 1.50 | 1.58 | 1.66 | 2.01 |

* Reference [22]

Table II. The calculated total and partial magnetic moments of Co(Fe)-doped MnNiSi compounds.

| Compounds | $M_{total}$ ($\mu_B$) | $M_{Mn}$ ($\mu_B$) | $M_{Co}$ ($\mu_B$) | $M_{Fe}$ ($\mu_B$) | $M_{Ni}$ ($\mu_B$) | $M_{Si}$ ($\mu_B$) |
|---|---|---|---|---|---|---|
| MnNiSi | 2.70 | 2.46 | | | 0.22 | 0.02 |
| Mn$_{0.75}$Co$_{0.25}$NiSi | 2.17 | 2.47 | 0.56 | | 0.17 | 0.01 |
| Mn$_{0.5}$Co$_{0.5}$NiSi | 1.63 | 2.52 | 0.46 | | 0.14 | 0 |
| Mn$_{0.75}$Fe$_{0.25}$NiSi | 2.41 | 2.36 | | 1.60 | 0.22 | 0.02 |
| Mn$_{0.5}$Fe$_{0.5}$NiSi | 2.21 | 2.47 | | 1.59 | 0.17 | 0.02 |

Figures of Captions

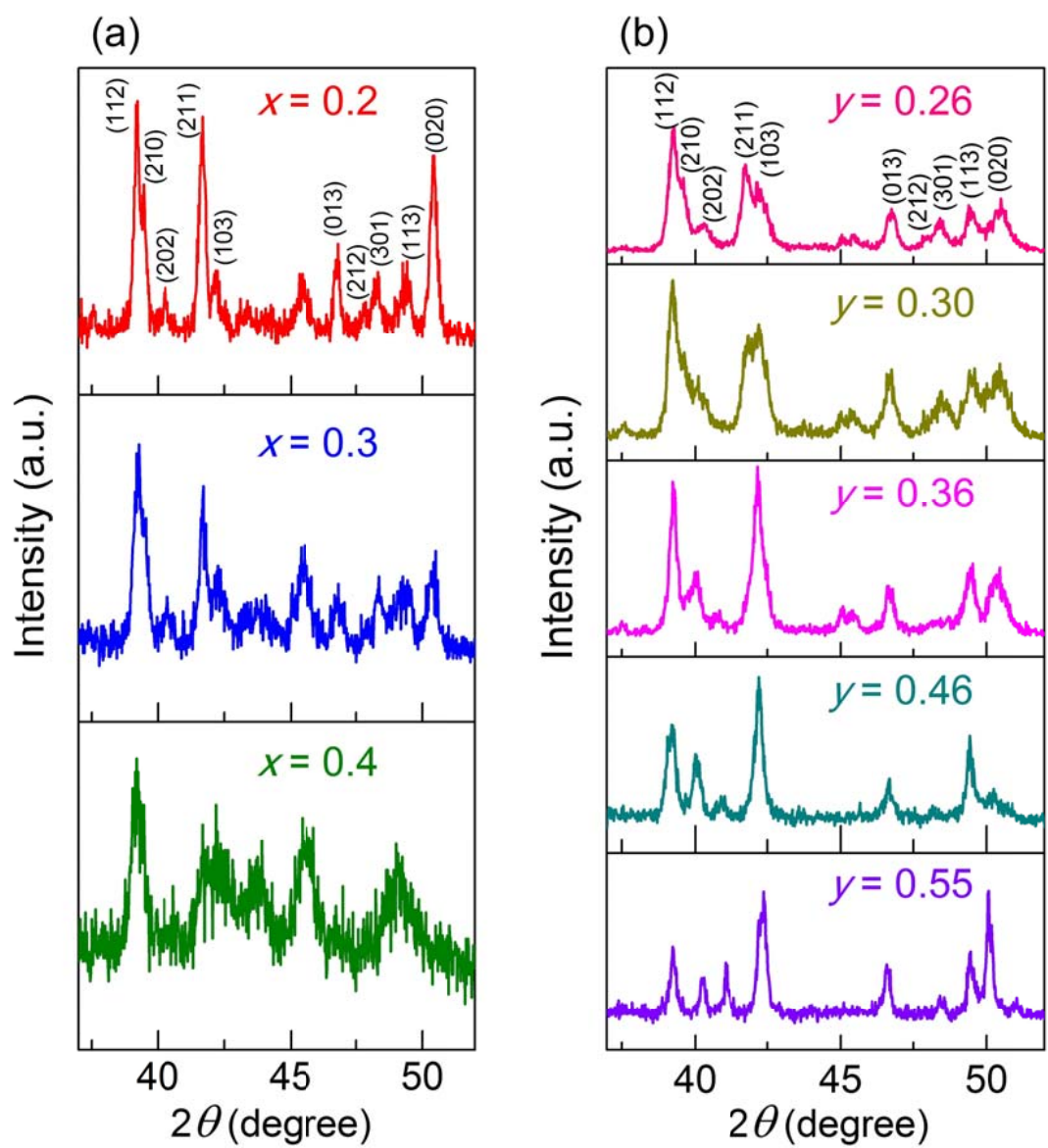

Fig. 1. Room-temperature XRD patterns of (a) $Mn_{1-x}Co_xNiSi$ ($x$ = 0.2, 0.3, 0.4) and (b) $Mn_{1-y}Fe_yNiSi$ ($y$ = 0.26, 0.30, 0.36, 0.46, 0.55). The TiNiSi-type orthorhombic structure for two systems was indexed.

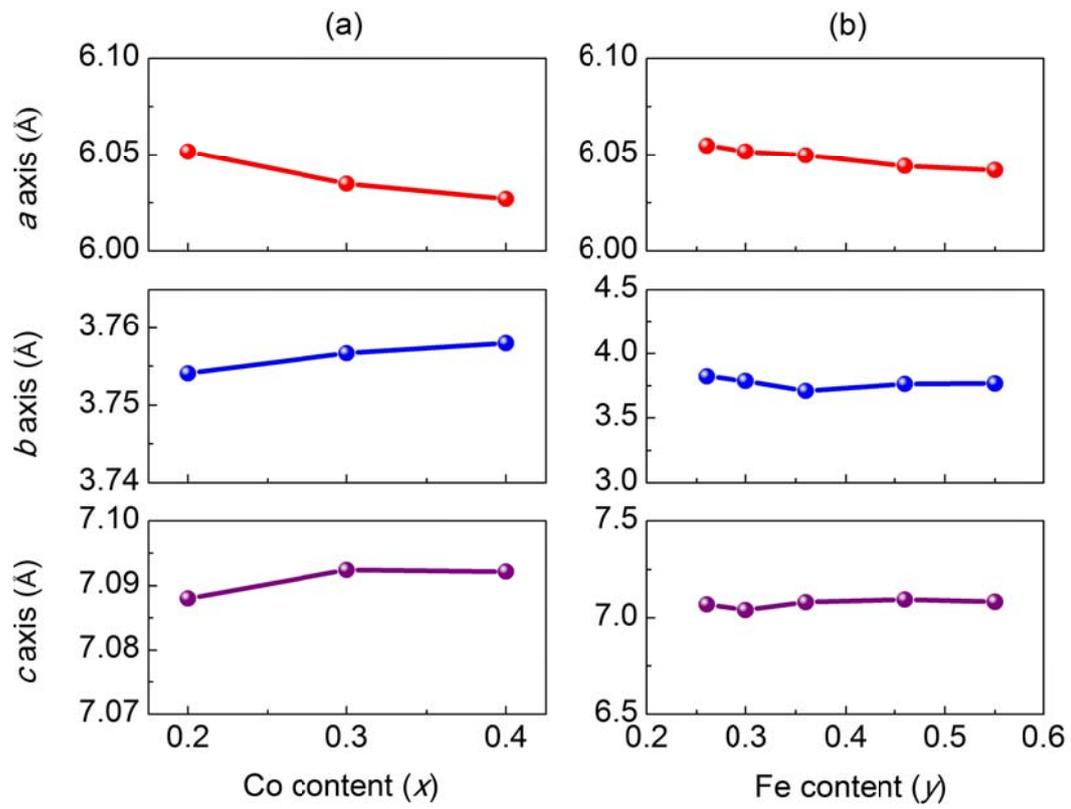

Fig. 2. The experimental lattice constants (unit: Å) *a*, *b* and *c* of (a) $Mn_{1-x}Co_xNiSi$ as a function of Co content and (b) $Mn_{1-y}Fe_yNiSi$ as a function of Fe content, respectively.

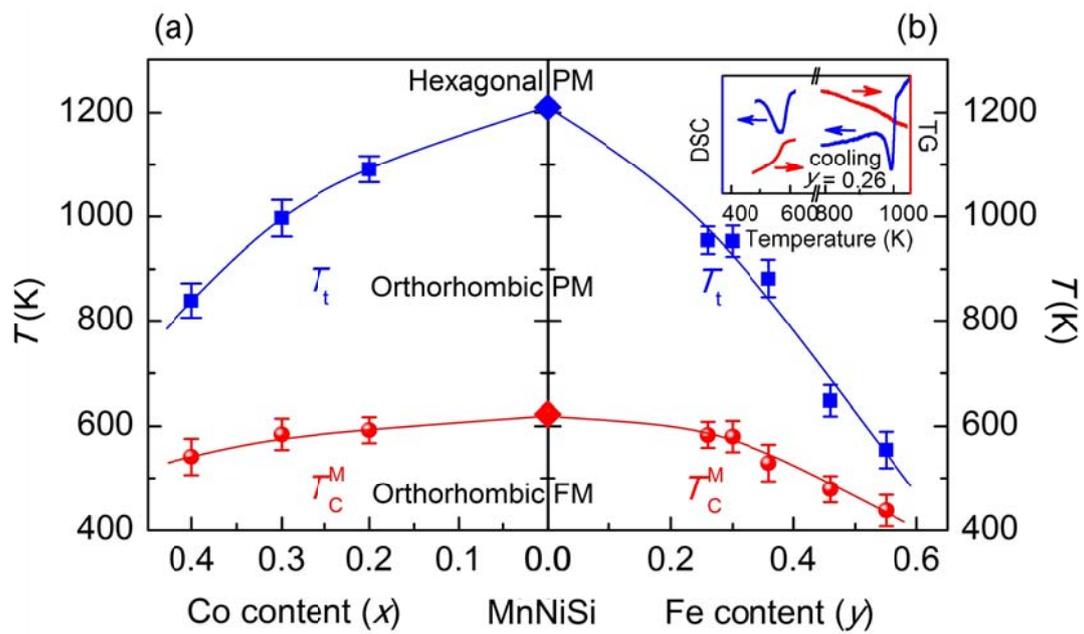

Fig. 3. Phase diagrams of (a) $Mn_{1-x}Co_xNiSi$ and (b) $Mn_{1-y}Fe_yNiSi$. The error bars are given by repeating the measurements.

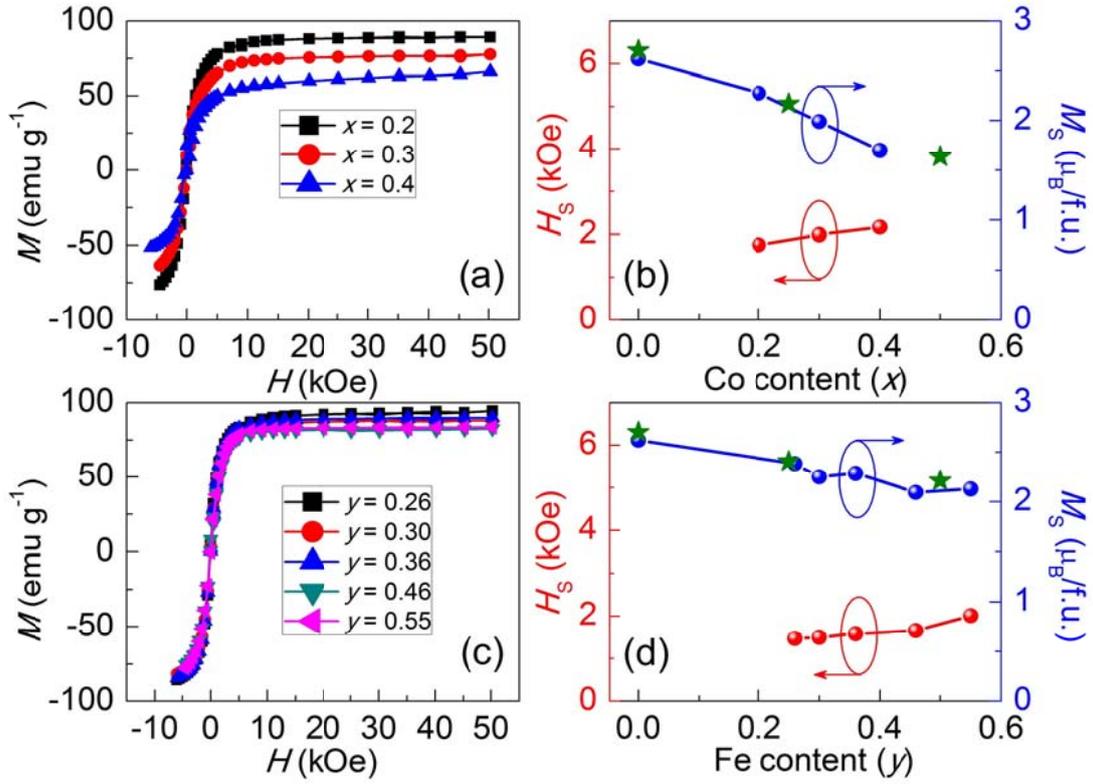

Fig. 4. (a) Magnetization curves and (b) Co-content dependence of saturation magnetization ($M_S$) and saturation field ($H_S$) for $Mn_{1-x}Co_xNiSi$ ($x$ = 0.2, 0.3, 0.4) samples measured at 5 K. (c) Magnetization curves and (d) Fe-content dependence of saturation magnetization ($M_S$) and saturation field ($H_S$) for $Mn_{1-y}Fe_yNiSi$ ($y$ = 0.26, 0.30, 0.36, 0.46, 0.55) samples measured at 5 K. Green pentagrams (★) in (b) and (d) represent the total magnetic moments calculated from first principles.

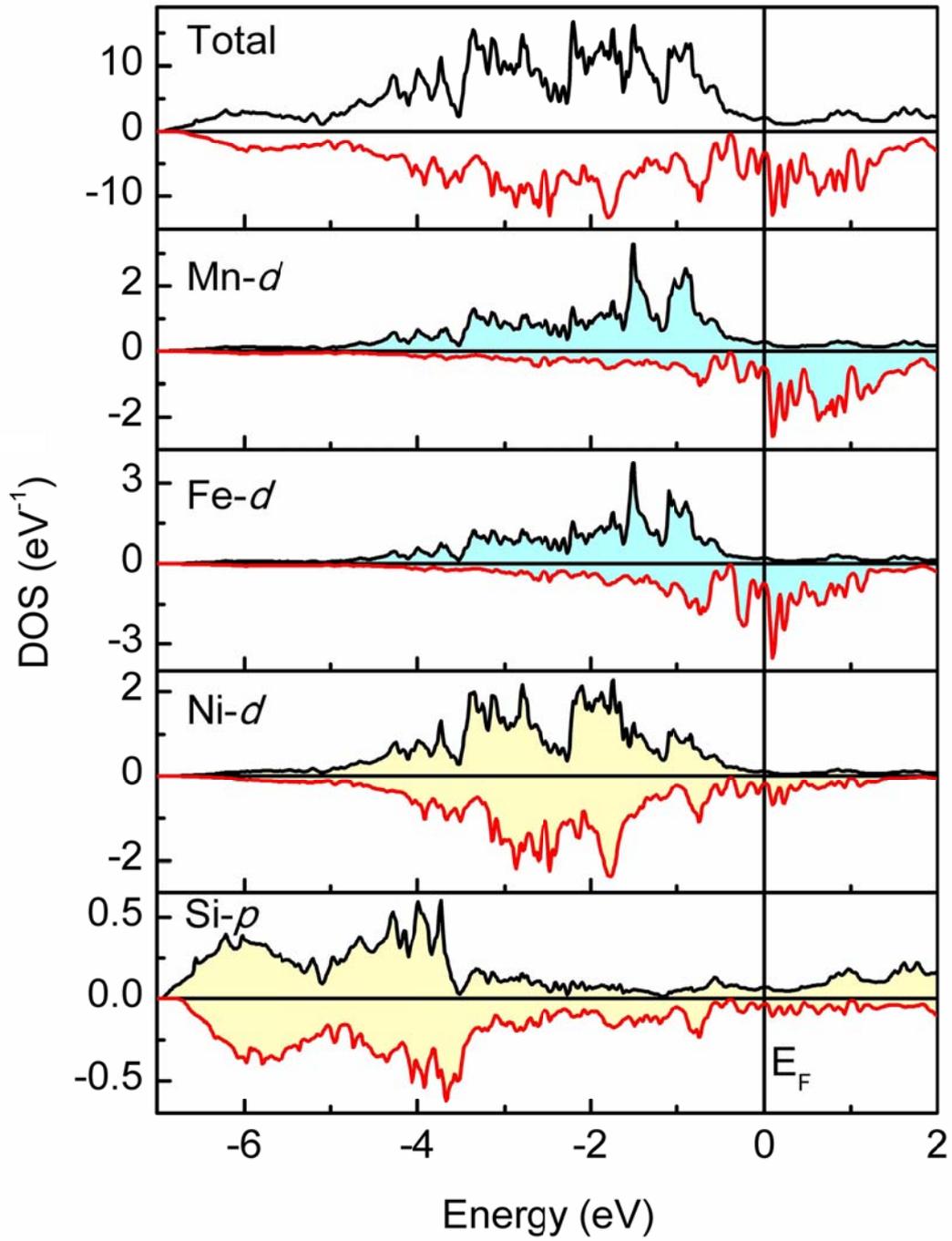

Fig. 5. Calculated total and partial density of state (DOS) of $Mn_{0.75}Fe_{0.25}NiSi$. Only *d*-electron DOS and *p*-electron DOS are given for 3*d* and *p*-block elements, respectively.